\documentclass[%
 reprint,
superscriptaddress,
nofootinbib,
 amsmath,amssymb,
 aps,
]{revtex4-1}

\usepackage[colorlinks,citecolor=blue]{hyperref}
\usepackage{graphicx}
\usepackage{dcolumn}
\usepackage{bm}
\usepackage{color}
\usepackage{placeins}
\usepackage{soul}

\definecolor{red}{rgb}{0.9, 0,0}
\definecolor{cerulean}{rgb}{0., 0.62,0.9}
\definecolor{navy}{rgb}{0.05, 0.05,0.8}

\begin{document}

\title{Searching for exotic B meson decays with the CMS L1 track trigger
}

\author{Jared A.~Evans}
\affiliation{Department of Physics, University of Cincinnati, OH 45221, USA}
\author{Abhijith Gandrakota}
\affiliation{Department of Physics and Astronomy, Rutgers University, Piscataway, NJ 08854, USA}
\author{Simon Knapen}
\affiliation{School of Natural Sciences, Institute for Advanced Study, Princeton, NJ 08540, USA}
\affiliation{CERN, Theoretical Physics Department, Geneva, Switzerland}
\author{Hardik Routray }
\affiliation{Department of Physics and Astronomy, Rutgers University, Piscataway, NJ 08854, USA}

\date{\today}

\begin{abstract}
The CMS phase II track trigger may allow for a displaced dimuon vertex trigger with
qualitatively lower $p_T$ thresholds than existing dimuon triggers. 
With this technique, we show that the CMS reach 
for exotic $B$-meson decays involving a displaced dimuon resonance, 
such as a light, Higgs-mixed scalar,
can be competitive with that of LHCb and Belle II. 
\end{abstract}

\maketitle


\section{Introduction}
The high luminosity runs of the LHC will deliver an enormous sample of $\sim10^{15}$ $B$ mesons to both ATLAS and CMS, far exceeding the number attainable at any other experiment in the near future. 
Capitalizing on this scientific opportunity will require the experiments to circumvent the trigger and background limitations inherent to a high energy hadron collider. The LHCb collaboration will do so through high precision tracking in the forward region, sophisticated online event reconstruction, and an order of magnitude reduction in instantaneous luminosity relative to ATLAS and CMS.
The phase II upgrades of ATLAS and CMS on the other hand will enable tracking at the hardware trigger \cite{Collaboration:2257755,collaboration:2714892}, which may substantially enhance their sensitivity to exotic $B$ decays \cite{collaboration:2714892,Gershtein:2019dhy}.

Concretely, it was shown that the CMS L1 track trigger could conceivably be configured to enable the reconstruction of displaced tracks with impact parameters as large as a few cm \cite{Gershtein:2017tsv,CMS:2018qgk,collaboration:2714892}. This opens up qualitatively new opportunities to trigger on signatures involving displaced jets \cite{Gershtein:2017tsv,CMS:2018qgk,Bhattacherjee:2020nno,collaboration:2714892,inprogress}. The L1 tracks can moreover be matched to the muon chamber \cite{collaboration:2714892}, and it was argued that as a result the background rate for a low threshold, dimuon vertex trigger may also not be prohibitive \cite{Gershtein:2019dhy}. 

 In this letter, we explore the {off-line} discovery potential of such a trigger in the context of exotic $B$-meson decays, with the aim of further motivating the experimental developments in this direction. In particular, we assert that the theory motivation for this program is very strong: A number of models feature a new dimuon resonance below the $B$ mass (see e.g.~\cite{Dev:2016vle,Burdman:2006tz,Chacko:2005pe,Bezrukov:2009yw,Martin:2014sxa,Krnjaic:2015mbs}), which is likely to be long-lived, providing an excellent handle to reject backgrounds.
For example, one of the most minimal extensions to the Standard Model (SM) adds a singlet scalar field $\phi$ which mixes with the SM Higgs through either
\begin{equation}\label{eq:lag}
\begin{split}
\mathcal{L}&\supset m\, \phi H^\dagger H\quad\text{with} \quad m \ll m_W, \text{ or}\\
\mathcal{L}&\supset -\mu^2\, \phi^2 +\epsilon \, \phi^2 H^\dagger H\quad\text{with} \quad \mu^2 > 0.
\end{split}
\end{equation}
In either case, in the mass basis, $\phi$ couples to all SM fermions proportional to their masses, but suppressed by the mixing angle $(s_\theta)$ between $\phi$ and the SM Higgs. For $m_K-m_\pi<m_\phi< m_B-m_K$, the dominant production mode at a hadron collider is through an electroweak penguin inducing the $B\to \phi X_s$ decay \cite{PhysRevD.26.3287,Chivukula:1988lo,Grinstein:1988yu}. Due to the extremely small width of the $B$ meson, this branching ratio can be large even for $s_\theta\ll1$. The branching ratio for $\phi \to\mu\mu$ is moreover between 0.1 and 0.01 for most of the relevant mass range \cite{Winkler:2018qyg}, although it is subject to substantial theoretical uncertainties. Importantly, existing limits on $s_\theta$ \cite{Aaij:2016qsm,Aaij:2015tna} already bound the minimal allowed lifetime of $\phi$ to be of the order of $\sim$ 1 cm. Displaced searches are therefore vital to probe this model any further. 

The LHCb collaboration has already performed dedicated searches for this model in the exclusive $B^+\to K^+ \mu^+\mu^-$ \cite{Aaij:2016qsm} and $B^0\to K^\ast \mu^+\mu^-$ \cite{Aaij:2015tna} channels, and has robustly excluded lifetimes $c\tau\lesssim 1$ cm for $m_\phi\lesssim 3$ GeV. Our proof-of-concept analysis differs from the LHCb approach in two crucial points: (i) we suggest an \emph{inclusive} search, which increases the signal acceptance, and (ii) the background will instead be suppressed by imposing isolation requirements and a hard cut on the transverse displacement of $\phi$'s decay vertex. 
Though it requires higher $p_T$ thresholds than LHCb, the geometric acceptance of a cylindrical detector such as CMS is substantially higher than that of LHCb's Vertex Locator for transverse displacements of more than a few cm.%
\footnote{For dark photon models on the other hand, $c\tau$ tends to be smaller and production is more forward, allowing substantial progress to be expected at LHCb in run 3 and beyond \cite{Ilten:2015hya,Ilten:2016tkc,Aaij:2017rft}. We suspect that in particular the exclusive approach of LHCb \cite{Ilten:2015hya} will outperform an inclusive strategy at CMS, but a detailed study of this scenario is left for future work. } 
When the additional selections imposed in the LHCb exclusive searches are also considered, we find a comparable signal efficiency in both cases, giving CMS an advantage due to its higher overall integrated luminosity.

This paper is organized as follows: In Sec.~\ref{sec:signal} we define the signal model and the corresponding event generation.  The main backgrounds are discussed in Sec.~\ref{sec:background} and the analysis strategy is presented in Sec~\ref{sec:analysis}. We present our results in Sec.~\ref{sec:results}.

\section{Signal definition\label{sec:signal}}

For the model in \eqref{eq:lag}, the production of the exotic state $\phi$ occurs through an electroweak penguin, with an estimated inclusive branching ratio of \mbox{$\text{Br}[B\to X_s \phi]\approx 6.2 \times s_\theta^2$} \cite{PhysRevD.26.3287,Chivukula:1988lo,Grinstein:1988yu}. One of the main features of an inclusive analysis at CMS is that it would be relatively insensitive to the particular decay mode of the $B^0/B^\pm$, since only the daughter muons of $\phi$ are being used. There is however a mild dependence on the exclusive decay channels of  $B^0/B^\pm$ due to the isolation criteria on the muons. For this reason, we implement the most important exclusive branching ratios \cite{Boiarska:2019jym} in our simulation of the signal (See Tab.~\ref{tab:branchingratios}). For the total branching ratio of $B^\pm/B^0$ into $\phi$ we will conservatively use the sum of the exclusive modes, which is about a factor of two smaller than the inclusive calculation. For the differential distributions we rely on \mbox{Pythia 8} \cite{Sjostrand:2006za}, which we normalize to the overall inclusive cross section as computed with FONLL \cite{Cacciari:1998it,Cacciari:2001td,Cacciari:2012ny,Cacciari:2015fta}. 

The decays of $\phi$ are determined by its effective Yukawa couplings to the lower generations of the SM fermion as well as through mixing with (broad) QCD spin 0 resonances. (See \cite{Winkler:2018qyg} for a recent calculation.)  The theoretical uncertainties on the lifetime of $\phi$ and branching fraction into muons are substantial, however these uncertainties will ultimately drop out from the projected limits we obtain in the $s_\theta$ vs $m_\phi$ plane, as explained in Sec.~\ref{sec:results}. While the model is fully specified by the mass and the mixing angle $(m_\phi, s_\theta)$, we will also present the result in a more model independent fashion, as is customary in the experimental literature. Concretely, the reach will be parametrized in terms of $m_\phi$,  $\text{Br}[B\to X_s \phi]\times\text{Br}[\phi\to\mu\mu]$ and $c\tau$, where the latter is the proper lifetime of $\phi$. 

It is very computationally inefficient to compute the signal acceptance by generating separate samples for each different value of $c\tau$. Instead, we generate samples with a \emph{stable} $\phi$ and analytically compute the weight for each event by evaluating
\begin{equation}
w_{c\tau}  =\int_{L_{xy}^-}^{L^+_{xy}}\!\!\! d L_{xy}\;  \epsilon(L_{xy}) e^{\frac{-L_{xy} \cosh\eta_\phi}{\beta\gamma c\tau}}.
\end{equation}
with $L_{xy}$ the distance of the vertex to the beamline.  $\eta_\phi$ and $\beta\gamma$ are the pseudorapidity and the boost of $\phi$. ${L_{xy}^-}$ and ${L^+_{xy}}$ represent the boundaries of the fiducial region under consideration and $\epsilon(L_{xy})$ is the estimated trigger efficiency from the Appendix of \cite{Gershtein:2019dhy}, which is above 0.8 over the region of interest.

\renewcommand{\arraystretch}{1.1}
\begin{table}[t]\centering
\begin{tabular}{l>{\centering\arraybackslash}p{2.5cm}>{\centering\arraybackslash}p{2.5cm}}
Channel &$m_\phi=0.5$  GeV& $m_\phi=2$  GeV\\\hline\hline
$B^\pm\to \phi K_1^\pm$&0.94&0.86\\
$B^\pm\to \phi K^{\ast\pm}_0$&0.86&0.97\\
$B^\pm\to \phi K^{\ast\pm}$&0.81&0.73\\
$B^\pm\to \phi K^\pm$&0.43&0.47\\
$B^\pm\to \phi K_2^{\ast\pm}$&0.29&0.11\\
$B^\pm\to \phi \pi^{\pm}$&0.012&0.014\\\hline
Total&3.35&3.16\\\hline\hline
$\phi\to\mu\mu$&0.12&0.18\\
\end{tabular}
\caption{ $\text{Br}[B^\pm\to \phi X_s ]/s_\theta^2$ for the dominant exclusive channels \cite{Boiarska:2019jym}. The branching ratios for various $K^\ast$, $K_1$ and $K_0^\ast$ resonances were summed together. Analogous branching ratios for $B^0$ can be obtained by multiplying with 0.93.  The last line shows $\text{Br}[\phi\to\mu\mu]$ as computed in \cite{Winkler:2018qyg}. \label{tab:branchingratios}}
\end{table}

\section{Backgrounds\label{sec:background}}

There are many potential backgrounds that can mimic the displaced dimuon signal, which can be usefully classified into two categories: non-$B$ and $B$ backgrounds.  Non-$B$ backgrounds include overlapping pileup muons, fake vertices, cosmic muons, and secondary vertices from interactions with the detector material. In what follows we assume these non-$B$ backgrounds can be removed with techniques such as masking known detector material, veto-ing muon pairs with large opening angles, etc, 
that we further assume will have a negligible impact on the signal efficiency.  These assumptions are perhaps bold, but are supported by the many ingenious searches for long-lived particles that have already been performed.

The $B$ backgrounds are all dimuon signatures that originate from a parent $b$ quark.  The displacement in these scenarios is mostly, if not entirely, due to the finite lifetime of the $B$ meson itself, with $c \tau_B \sim 500 \mu$m.  A hard cut on the displacement of the secondary vertex implies that the $B$ mesons which do decay in the signal region are typically very boosted. The daughter muons of the $B$ are therefore usually not isolated. The $B$ backgrounds include $b\to X_s \mu^+\mu^-$, $b \to X + \psi(nS)\to  \mu^+\mu^-$, and   $b\to \mu \nu (X_c \to \mu \nu X_s)$.  We will discuss these backgrounds in the following paragraphs.

The BR$(b\to X_s\mu^+\mu^-) \sim  4\times 10^{-6}$ is small enough to suppress it far below the other two major $B$ backgrounds.  The $J/\psi(1S)$ resonance on the other hand appears in over 1\% of all $B$ meson decays, and nearly 6\% of those decay into $\mu^+\mu^-$.   This large background, along with other resonances, including the $\eta$, $\rho$, $\omega$, $\phi(1020)$, and  higher $\psi(nS)$,  are confined the narrow mass peak and could be veto-ed easily if needed. 
However, the cuts we will apply in Sec.~\ref{sec:analysis} seem effective at controlling even the $J/\psi$ background without vetoing the mass window. 

The largest contribution to the continuum of the dimuon invariant mass distribution is from events where the $B$ undergoes a semi-leptonic decay, $b\to \mu \nu X_c$, followed by a semi-leptonic $D$ decay providing the second muon.  As each semi-leptonic branching fraction is a little over 10\% to muons, na\"ively $10^{13}$ $B$s at the HL-LHC would result in this signature.   However, there are some useful ways to reduce this background.  First, due to the secondary displacement of the $D$ meson, the muon tracks from the $B$ and $D$ will not form a vertex at truth level. While it is often the case that the resulting vertexing fit is of high enough quality to mimic a common point of origin, a sizable fraction can be discarded based on a vertexing requirement.  We impose a conservative vertexing quality cut of $\sigma_v =100\,\mu$m, i.e., that the distance of closest approach for the two tracks is less than $\sigma_v$,\footnote{For all our background samples, we conservatively assume 100\% reconstruction and trigger efficiency for both muons, regardless of their impact parameters.}  which rejects $\sim$50\% of the background. Second, the $p_T$ of the two muons are typically fairly asymmetric, which is not the case for the signal. In practice this means that the muon from the $D$ meson decay often falls below our $p_T$ threshold.

\begin{figure}[t]
\includegraphics[width=0.45\textwidth]{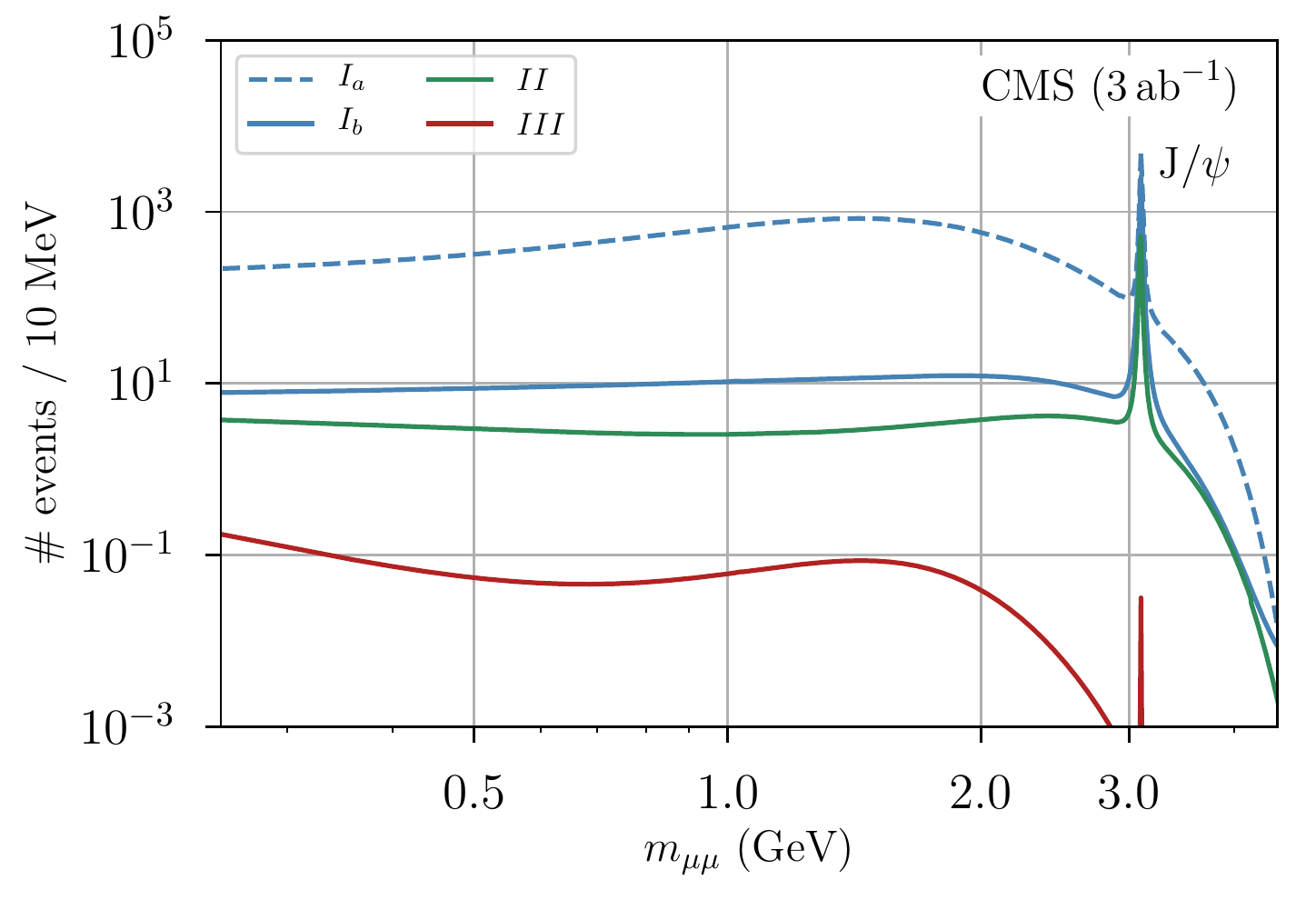}
\caption{Background component from $B$-meson decays for the cuts specified in Tab.~\ref{tab:cutflow}. (See Sec.~\ref{sec:analysis} for details.)\label{fig:continuum}}
\end{figure}

The $B$ background differential distributions where simulated with Pythia 8 \cite{Sjostrand:2006za}, and its overall cross section normalized to $500 \mu$b. Fig.~\ref{fig:continuum} shows the resulting invariant mass spectrum, subjected to the various cuts outlined in Sec.~\ref{sec:analysis}. (See Tab.~\ref{tab:cutflow} for a summary.) The $J/\psi$ resonance is clearly visible. The $\omega, \rho,\phi$ and $\eta$ mesons were also explicitly included, but their contribution after cuts proved to be negligible compared to the continuum. The solid (dashed) blue curves assume only a few baseline cuts with (without) imposing an isolation criterion on both muons. The green and red curves indicate the effects of additional cuts on the $\phi$'s impact parameter and the muon $p_T$ respectively, as detailed in the next section.

\section{Analysis strategy\label{sec:analysis}}
Nearly all of the B background can be controlled by simply cutting very hard on the transverse distance between the decay vertex and the beamline ($L_{xy}$). In addition we suggest a few additional cuts which could further reduce the background by a few orders of magnitude, at minimal cost to the signal efficiency. The cut flow is summarized in Tab.~\ref{tab:cutflow}. The definition of our variables is explained in the text below.

The CMS L1 track trigger may reasonably record dimuon pairs with a $p_T$ as low as roughly 4 GeV each \cite{Gershtein:2019dhy}. To reduce uncertainties due to so far unknown trigger efficiency turn-on near threshold, we will require $p_T>4.5$ GeV. Both muons are moreover required to satisfy $|\eta|<2.4$. For the estimated efficiency of the trigger we follow \cite{Gershtein:2019dhy}, which is based on a simplified simulation of the L1 track trigger developed in \cite{Gershtein:2017tsv}. In particular, for $L_{xy}>35$ cm the efficiency drops to zero as the muons leave an insufficient number of stubs in the outer tracker to reliably reconstruct a track. We require $L_{xy}<30$ cm and take $L_{xy}>7.5$ cm as a baseline cut to suppress the $B$ background.\footnote{Alternatively, one may cut on the transverse impact parameter of the individual tracks ($|d_0|$), as was done for in the CMS search for displaced lepton jets \cite{CMS:2014hka}.}  

\begin{figure*}[t]\centering
\includegraphics[width=0.42\textwidth]{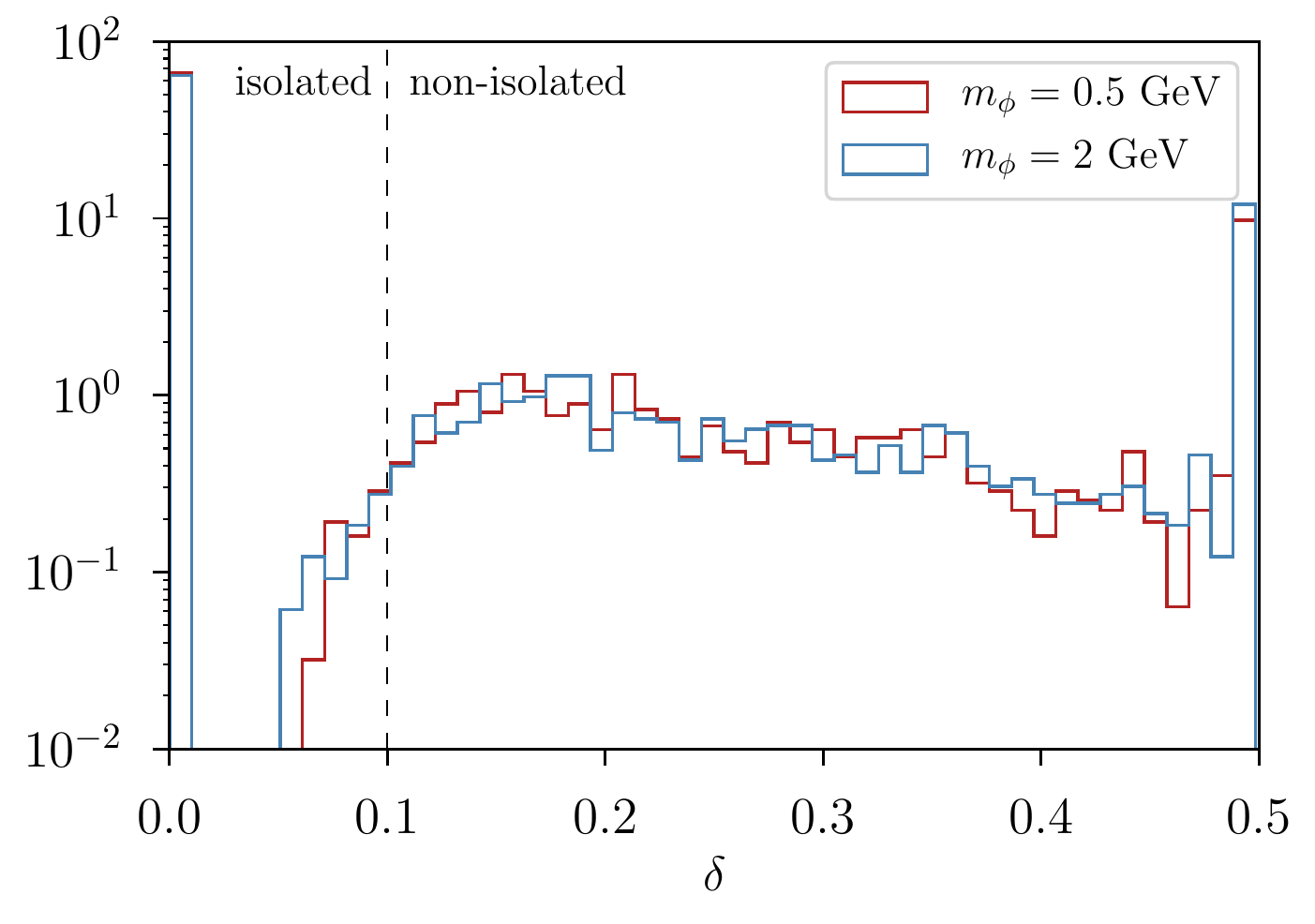}\hfill
\includegraphics[width=0.42\textwidth]{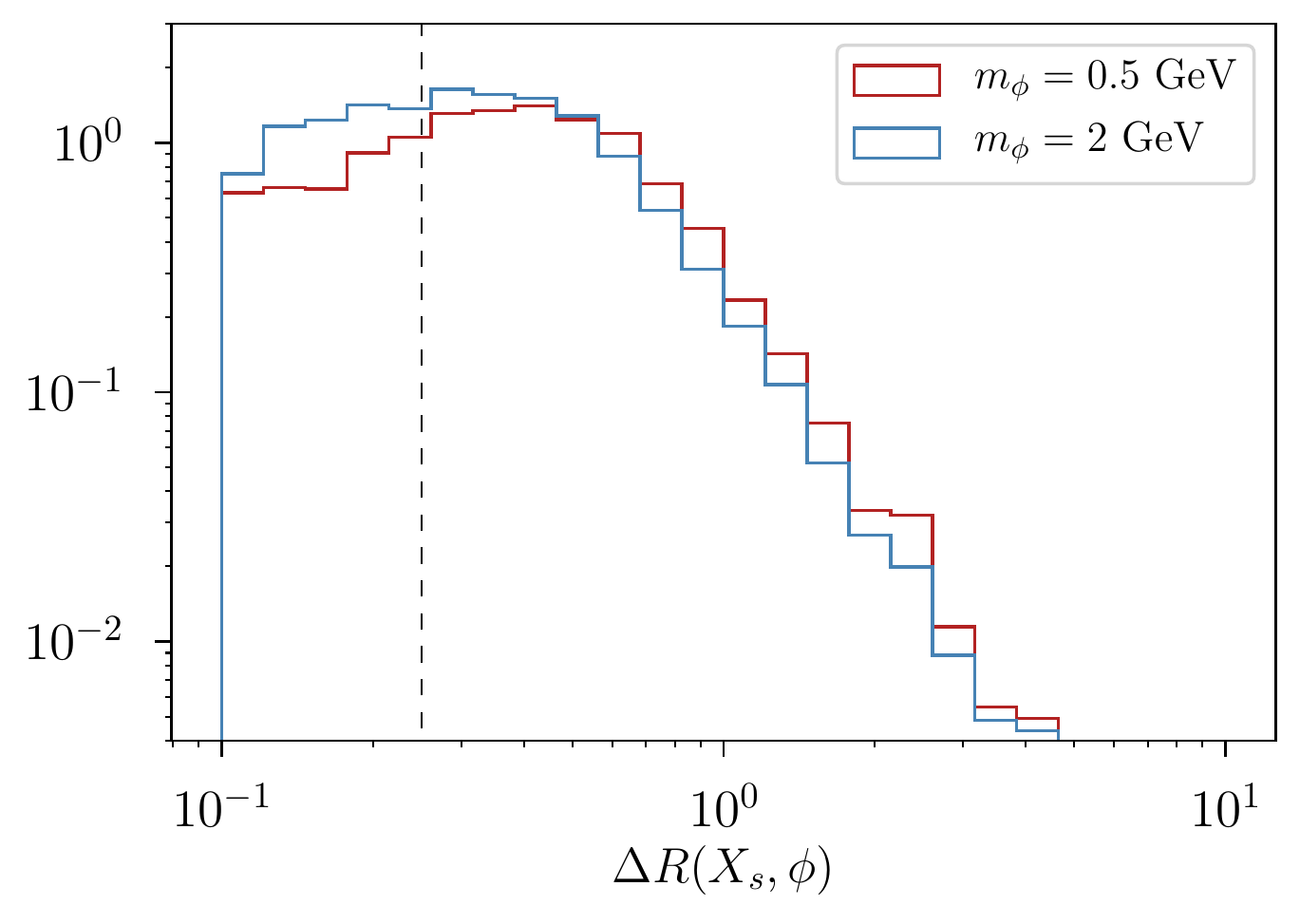}
\caption{\emph{Left:} Isolation variable $\delta$ for two signal benchmarks. The rightmost bin is an overflow bin, containing all events with $\delta>0.5$. The isolation efficiencies for the $m_\phi=$0.5 GeV and $m_\phi=2$ GeV benchmarks are 0.68 and 0.66 respectively. \emph{Right:} Truth-level $\Delta R$ between $\phi$ and its sister meson. The vertical dashed line indicates the isolation cone of $\Delta R=0.25$.\label{fig:isolation}
}
\end{figure*}

We further define an isolation variable ($\delta$) for each muon as the scalar sum of the $p_T$ of each track with $p_T>0.7$ GeV within a cone defined by $\Delta R<$ 0.25, divided by the muon $p_T$:
\begin{equation}
\delta \equiv \sum_{i\in \text{tracks}} \frac{p_{T_i}}{p_T{_{\mu}}}\quad\text{with}\quad\begin{array}{l} p_{T_i} >0.7 \;\mathrm{GeV}\\ \Delta R <0.25\end{array}
\end{equation}
The muons themselves are not counted towards each others isolation variable $\delta$. A muon is considered isolated if $\delta<0.1$; we require both muons to satisfy this criterion. Although the dimuon pair originates from a $B$ decay, both muons are nevertheless isolated in about  2 out of 3 events passing the other fiducial selections outlined above. Due to the mild $p_T$ requirements, the $B$ mesons which produce most of the signal only have a boost of $\mathcal{O}(\text{few})$, leading to relatively wide opening angles between $\phi$ and its strange sister meson. This is shown in the right-hand panel of Fig.~\ref{fig:isolation}, for the exclusive decay paths in Tab.~\ref{tab:branchingratios}. While this feature must be verified in data, e.g.~by making use of $B\to J/\psi\, X$ transitions, in simulation it is robust when initial and final state radiation are included.

\begin{figure*}[t!]\centering
\includegraphics[width=0.45\textwidth]{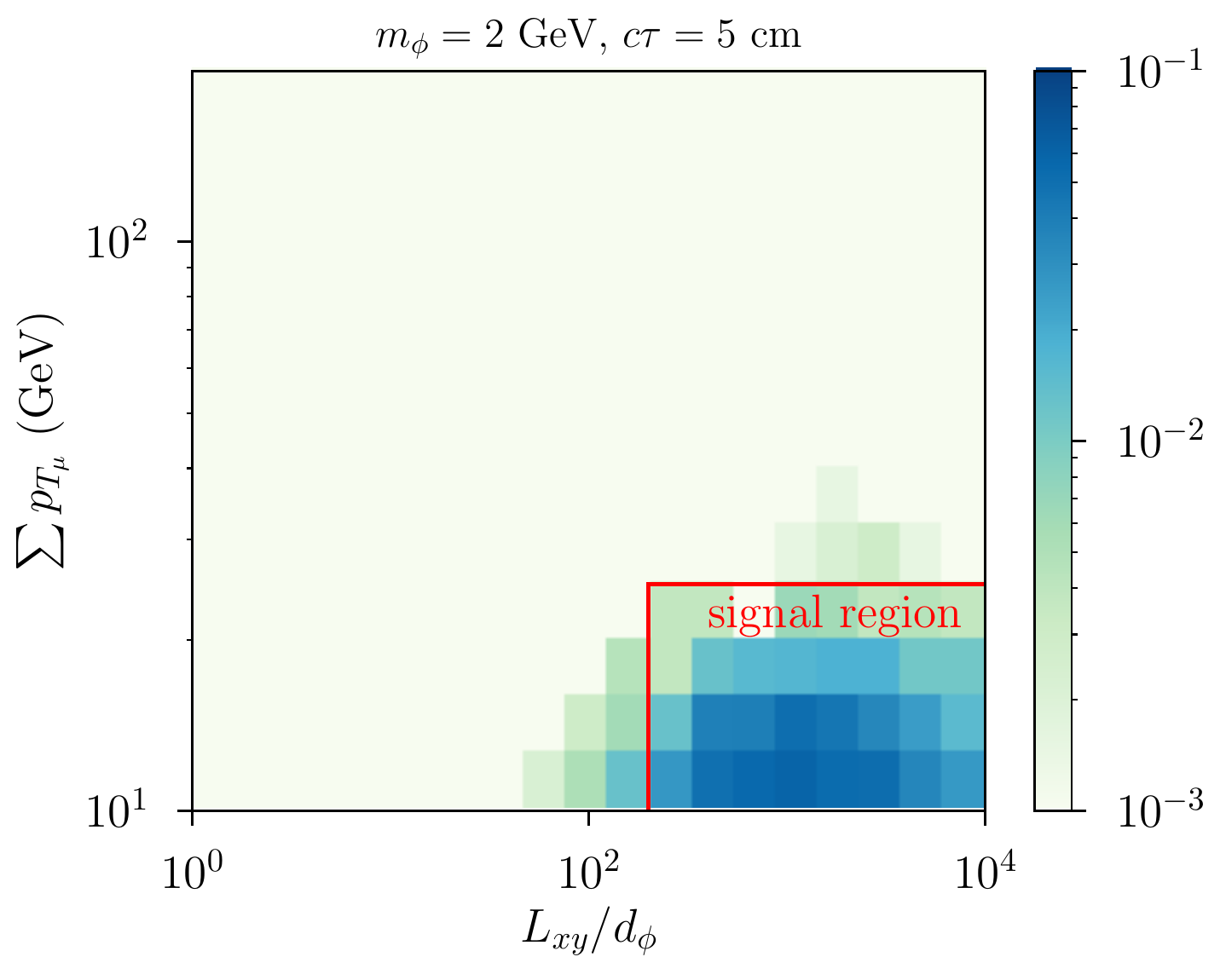}\hfill
\includegraphics[width=0.45\textwidth]{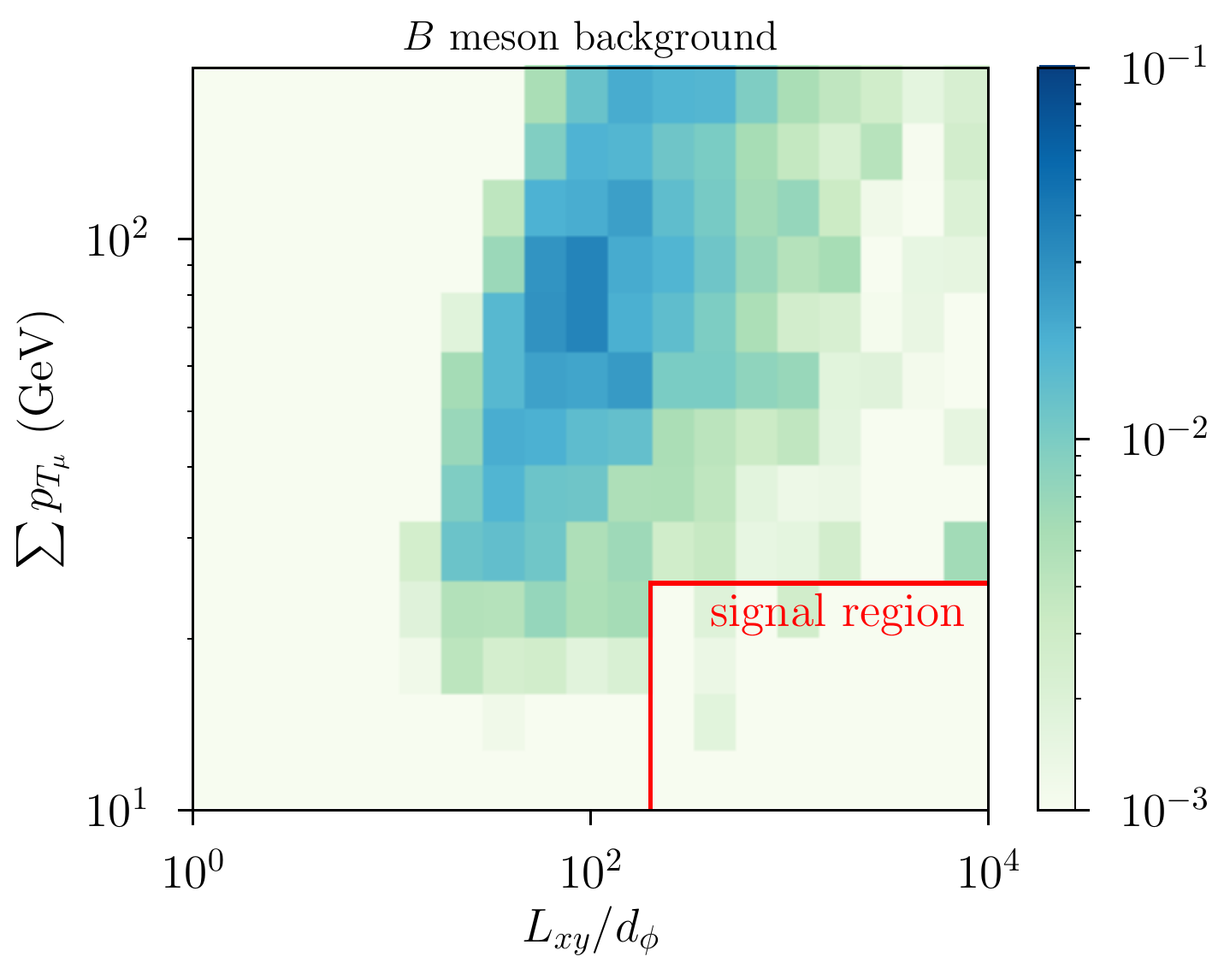}
\caption{Distribution of a signal benchmark (left) and background (right) for the $L_{xy}/d_\phi$ and $\sum p_{T_\mu}$ variables, after imposing the baseline cuts and isolation requirements ($I_b$ in Tab.~\ref{tab:cutflow}). The red box indicates the signal region.\label{fig:additionalcuts}
}
\end{figure*}

\begin{table}[t]\centering
\begin{tabular}{l>{\centering\arraybackslash}p{6.5cm}>{\centering\arraybackslash}p{1.5cm}}
 & cuts & signal eff. \\ \hline\hline
$I_a$ &$p_T>4.5$ GeV and $|\eta|<2.4$ and $L_{xy}>7.5$ cm&$2\times 10^{-3}$\\
$I_b$ &$I_a$ and $\delta<0.1$ &0.68\\
$II$ &$I_b$ and $L_{xy}/d_\phi>200$&0.96\\
$III$ &$II$ and $\sum p_{T_\mu} <25$ GeV&0.98\\
\end{tabular}
\caption{Efficiency of each consecutive set of cuts, relative to the preceding set or cuts, for $m_\phi=2$ GeV, $c\tau=5$ cm. $I_a$ and $I_b$ represent the baseline cuts without and with isolation respectively. The $I_a$ selection was normalized with respect to the inclusive $B$-meson cross section of 500 $\mu$b times $\text{Br}[B\to X_s \phi]\times\text{Br}[\phi\to\mu\mu]$. The cuts $II$ and $III$ in particular further reduce the background (See Fig.~\ref{fig:continuum}) without significantly impacting the signal efficiency.  \label{tab:cutflow}}
\end{table}

Aside from the simple baseline selection laid out above, we suggest two additional cuts which prove very effective at reducing the $B$ backgrounds. Firstly, one observes that in the limit where $c\tau_B\to0$ and perfect reconstruction, the reconstructed trajectory of $\phi$ should point back to the primary vertex. Defining the parameter $d_\phi$ as the distance of closest approach between the primary vertex and the reconstructed $\phi$ trajectory, we thus find the following scaling
\begin{equation*}
\begin{array}{c|cc}&\text{background}&\text{signal}\\\hline
L_{xy}&\sim c\tau_B & \sim c\tau_{\phi}\\
d_\phi&\sim c\tau_B & \sim c\tau_B
\end{array}
\end{equation*}
in the limit where $c\tau_\phi\gg c\tau_B$.
As shown by the green curve in Fig.~\ref{fig:continuum}, a hard cut on the dimensionless ratio $L_{xy}/d_\phi>200$ is therefore very effective at further suppressing he $B$-meson background, with only $\sim 4\%$ reduction in signal efficiency.

Finally, given that $c\tau_B\ll 7.5$ cm, any background events that that pass the above set of cuts are typically very boosted. Cutting on the scalar sum of the muon $p_T$, $\sum p_{T_\mu}<25$ GeV, removes any residual, boosted background events, without reducing the signal efficiency. The distributions of the $L_{xy}/d_\phi$ and  $\sum p_{T_\mu}$ variables and their correlations are shown in Fig.~\ref{fig:additionalcuts} for signal and background.  Due to the substantial anti-correlation of these two cuts on the background, they are most effective when applied together.

\section{Results\label{sec:results}}

\begin{figure*}[t]\centering
\includegraphics[width=0.45\textwidth]{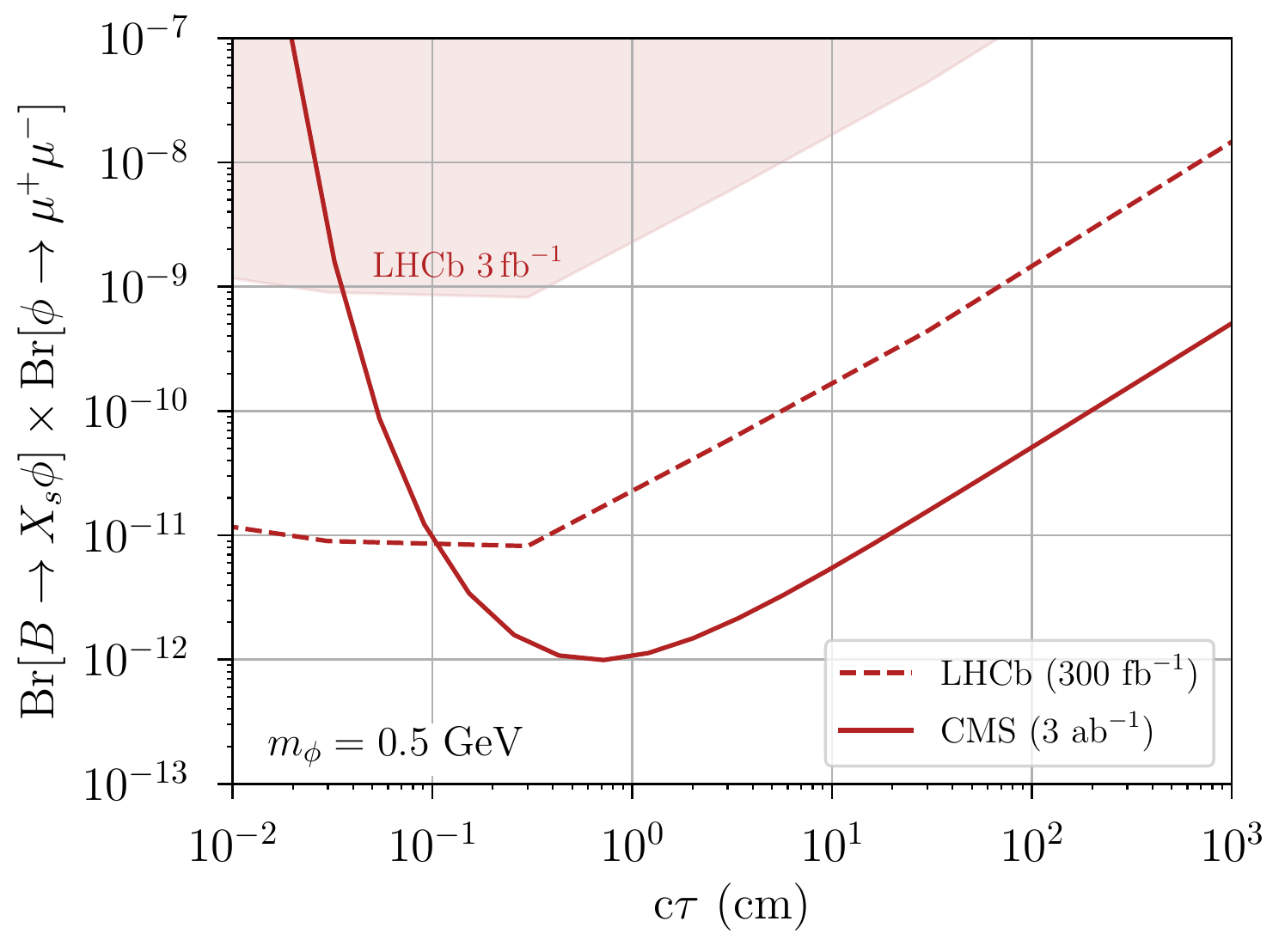}\hfill
\includegraphics[width=0.45\textwidth]{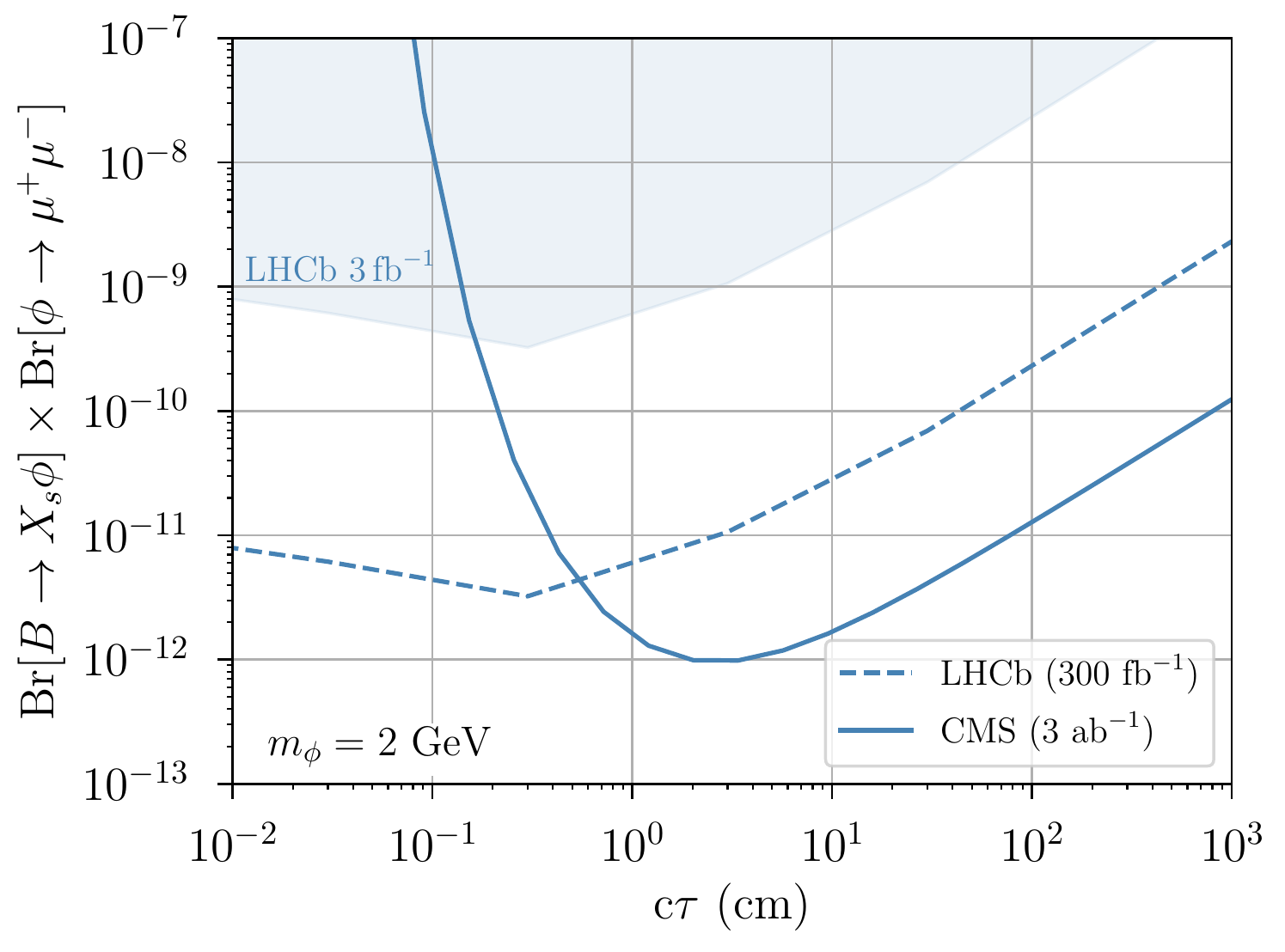}
\caption{Projected reach in the model independent parametrization, for the cuts in Tab.~\ref{tab:cutflow}. {\bf Left:} $m_\phi=0.5$ GeV;   {\bf Right:} $m_\phi=2$ GeV.  Shaded region and dashed line represent the existing LHCb limits \cite{Aaij:2016qsm} and an optimistic extrapolation of the LHCb reach. (See text.)\label{fig:moneyplots}
}
\end{figure*}

In our reach estimates, we impose the most stringent selection, corresponding to signal region III in Tab.~\ref{tab:cutflow}. This corresponds to the the red curve in Fig.~\ref{fig:continuum}, where we see that the $B$-meson background can be reduced to negligible levels, assuming an invariant mass resolution of $\sim 10$ MeV. The resulting sensitivity in the model-independent parametrization is show in Fig.~\ref{fig:moneyplots}. The CMS HL LHC data will improve the existing limits by several orders of magnitude, and could moreover outperform LHCb's HL LHC reach. For this comparison, we extrapolated the LHCb reach to 300 $\text{fb}^{-1}$ in the most optimistic manner, by assuming that the existing analysis continues to be largely background free, with negligible losses in signal efficiency. We reiterate however that our results for CMS are also optimistic, since we assumed that the non-$B$ backgrounds can be suppressed to negligible levels, which, while plausible, may not be attainable in practice.

In the regime where the lab frame lifetime of $\phi$ exceeds the lower cut on $L_{xy}$, the reach scales as $\sim c\tau/(L_{xy}^{+}-L_{xy}^{-})$ with $L_{xy}^{+}$ ($L_{xy}^{-}$) the upper (lower) cut on $L_{xy}$.  It is therefore natural to ask whether the $L_{xy}$ upper cut can be relaxed further by using stand-alone muons. Searches of this sort in fact already \cite{CMS-PAS-EXO-14-012,Aad:2019tua} exist, but due to the reduced momentum and vertex resolutions for stand-alone muons, we suspect that this type of analysis would have substantially larger backgrounds or would require higher $p_T$ thresholds. 

\begin{figure}[t]\centering
\includegraphics[width=0.48\textwidth]{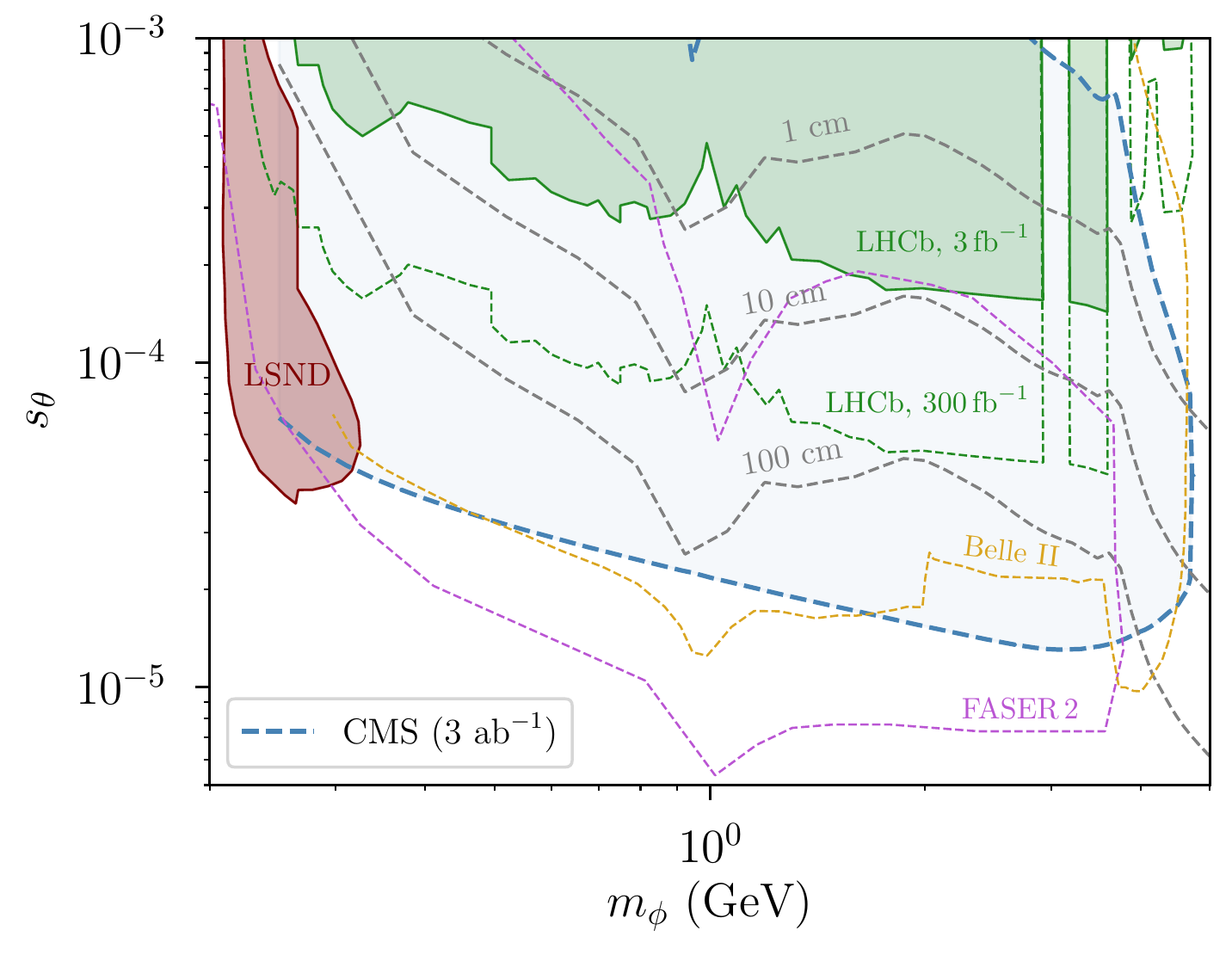}
\caption{Projected reach in terms of the mass ($m_\phi$) and mixing angle ($s_\theta$) of a light scalar mixing with the SM Higgs.  Shaded regions represent the existing LHCb \cite{Aaij:2016qsm} and LSND \cite{Foroughi-Abari:2020gju} limits. The dashed gray contours indicate the proper lifetime of $\phi$ \cite{Winkler:2018qyg}. We also show projections for Belle II \cite{Kachanovich:2020yhi} and the FASER upgrade \cite{Feng:2017vli}. \label{fig:moneyplotHiggs}
}
\end{figure}

In addition to the model-independent parametrization, it is useful to map the reach on the the concrete model of a light scalar mixing with the Higgs, as defined by Eq.~\ref{eq:lag}. The resulting reach is shown in Fig.~\ref{fig:moneyplotHiggs}. In addition to the existing bounds from LHCb \cite{Aaij:2016qsm} and LSND \cite{Foroughi-Abari:2020gju}, we furthermore show contours of the proper lifetime of $\phi$ in this model. In the lower edge of the reach, this reveals that the typical lab frame lifetime of $\phi$ exceeds the spacial dimensions of the fiducial volume. This implies the signal yield is proportional to 
\begin{align}
N_{sig}&\sim \text{Br}[B\to X_s \phi]\times\text{Br}[\phi\to\mu\mu] \times \frac{L_{xy}^{+}-L_{xy}^{-}}{c\tau}\times \mathcal{L}\nonumber\\
&\sim \frac{\Gamma_{\phi\to\mu\mu}}{s_\theta^2} \times (L_{xy}^{+}-L_{xy}^{-})\times s^{4}_{\theta}\times \mathcal{L}
\end{align}
where $\Gamma_{\phi\to\mu\mu}$ is the partial width of $\phi$ to muons and $\mathcal{L}$ is the integrated luminosity. Since the $\Gamma_{\phi\to\mu\mu}/s_\theta^2$ combination is independent of $s_\theta$, this makes explicit that the reach, when mapped on to ($m_\phi,s_\theta$), scales as the $4^\mathrm{th}$ root of the luminosity, under the assumption of zero background. In addition, the total width of $\phi$ drops from the expression, greatly reducing the theoretical uncertainties. This moreover explains why the projected limit is a relatively featureless curve, especially in comparison to the more complex lifetime contours. 

Looking ahead, both LHCb and Belle II are in strong positions to (further) weigh in on this signature. LHCb in particular is expected to continue to provide the best limits for $c\tau\lesssim 1\,\mathrm{cm}\times \frac{m_\phi}{\mathrm{GeV}}$, while Belle II is expected to further improve limits in the long lifetime regime, especially with their ultimate data set \cite{Filimonova:2019tuy,Kachanovich:2020yhi}. More broadly, low mass dimuon resonances may be produced in a variety of other ways, such as through exotic Higgs decays or as part of more elaborate hidden sectors. The HL LHC in general and CMS in particular could have excellent reach for such resonances, regardless how they are being produced.

\begin{acknowledgments}
We thank  Amit Lath for discussions leading up to this study and Nick Amin, Brian Batell, Claudio Campagnari, Matthew Citron, Yuri Gershtein, Vladimir Gligorov, Amit Lath, Zoltan Ligeti, Bennett Marsh, Dean Robinson, Ulascan Sarica and David Stuart for useful discussions throughout. The work of SK was supported by DOE grant DE-SC0009988.  This research used resources of the National Energy Research Scientific Computing Center (NERSC), a U.S. Department of Energy Office of Science User Facility operated under Contract No. DE-AC02-05CH11231. JAE is supported in part by the DOE grant DE-SC0011784.  This work was initiated at the Aspen Center for Physics, which is supported by National Science Foundation \mbox{grant PHY-1607611.}
\end{acknowledgments}

\FloatBarrier

\bibliographystyle{apsrev4-1}
\bibliography{CMS_HLLHC}

\end{document}